\documentclass{webofc}
\usepackage[varg]{txfonts}   
\DeclareMathOperator\arctanh{arctanh}
\usepackage{graphicx}
\usepackage{subcaption}
\usepackage{commath}
\usepackage[colorlinks,citecolor=blue,urlcolor=blue,bookmarks=false,hypertexnames=true]{hyperref} 
\usepackage{natbib}
\usepackage[colorinlistoftodos]{todonotes}

\begin{document}
\title{Multi-particle reconstruction in the High Granularity Calorimeter using object condensation and graph neural networks}

\author{\firstname{Shah Rukh} \lastname{Qasim}\inst{1,2}\fnsep\thanks{\email{shah.r.qasim@stu.mmu.ac.uk,  shah.rukh.qasim@cern.ch}} \and
        \firstname{Kenneth} \lastname{Long}\inst{1}\fnsep\thanks{\email{kenneth.long@cern.ch}} 
\and
        \firstname{Jan} \lastname{Kieseler}\inst{1}\fnsep\thanks{\email{jan.kieseler@cern.ch}}
\and
        \firstname{Maurizio} \lastname{Pierini}\inst{1}\fnsep\thanks{\email{maurizio.pierini@cern.ch}}
             for the CMS Collaboration, and
        \firstname{Raheel} \lastname{Nawaz}\inst{2}\fnsep\thanks{\email{R.Nawaz@mmu.ac.uk}}
}

\institute{CERN, EP/CMG 
\and  Manchester Metropolitan University
          }

\abstract{%
The high-luminosity upgrade of the LHC will come with unprecedented physics and computing challenges. One of these challenges is the accurate reconstruction of particles in events with up to 200 simultaneous proton-proton interactions. The planned CMS High Granularity Calorimeter offers fine spatial resolution for this purpose, with more than 6 million channels, but also poses unique challenges to reconstruction algorithms aiming to reconstruct individual particle showers. In this contribution, we propose an end-to-end machine-learning method that performs clustering, classification, and energy and position regression in one step while staying within memory and computational constraints. We employ GravNet, a graph neural network, and an object condensation loss function to achieve this task. Additionally, we propose a method to relate truth showers to reconstructed showers by maximising the energy weighted intersection over union using maximal weight matching. Our results show the efficiency of our method and highlight a promising research direction to be investigated further.
}
\maketitle
%


\section{Introduction}
\label{intro}
The high-luminosity upgrade of the LHC (HL-LHC) will pose unprecedented computational challenges. In order to reach the target luminosity, collisions at the HL-LHC are planned to have an average of 200 proton-proton interactions per bunch crossing~\cite{Apollinari:2284929}. 
Building tracks and energy clusters and associating them to particles of the primary hard interaction, or to additional interactions in the collision (pileup), in such a complex environment is a highly challenging task that requires the full event to be reconstructed in the most detailed way before an assignment can be made. 

The physics program of the HL-LHC seeks to exploit an expected 3000 fb$^{-1}$ of pp collisions delivered by the end of the HL-LHC project. Thanks to this large data set, it will be possible to characterize the electroweak sector of the standard model with unprecedented precision. Vector boson scattering and vector boson fusion (VBF) processes, where vector bosons are radiated from the incoming quarks of the colliding protons before directly interacting, are important probes of such interactions. A distinct feature of VBF processes is high-momentum jets in the forward part of the detector. Consequently, identifying these processes and accurately reconstructing their properties requires accurate reconstruction in the very particle-dense forward region of the detector. To enable this, and to cope with the high radiation environments in the forward region at high pileup, the calorimeter endcaps of the CMS detector \cite{collaboration2008cms} will be replaced for the HL-LHC operations with the CMS High Granularity Calorimeter (HGCAL), providing both fine lateral and longitudinal segmentation~\cite{CERN-LHCC-2017-023}.

The HGCAL detector is a sampling endcap calorimeter, comprising 50 sensor and absorber layers with a total thickness of about 10 hadronic interaction lengths ($\lambda$). The first 28 layers correspond to about 25 radiation lengths or 1.3 $\lambda$, consist of silicon sensors and absorber material and form the electromagnetic section. The sensors are hexagonal in shape and have thicknesses of 120, 200, or 300$\,\mathrm \mu$m, depending on the expected fluence. Their size ranges from about 0.5 to 1.2$\,\mathrm{cm^2}$, with higher granularity closer to the beam pipe. The following hadronic section includes 12 fine sampling layers. The remaining layers have a larger fraction of absorber material. Also here, the sensor size increases with distance to the beam pipe. In regions of lower expected fluences, and therefore increasing distance to the beam pipe and the interaction point, the silicon sensors are replaced by scintillator tiles, regular in $\eta$ and $\phi$ and equipped with silicon photomultipliers.

Local reconstruction of particle tracks and clusters in the HGCAL is agnostic to the full event record from the CMS detector. In particular, without the association of calorimeter tracks and clusters to the tracking detector, assigning an object to the initiating vertex, and thus to the primary interaction or to pileup, is not practical at this stage. It is therefore crucial to be able to identify and determine the properties of each individual shower down to low energy, even in case of severe overlaps with other incident particles. For this reason, it might be suboptimal to adopt a binary or sequential reconstruction approach, in which an attempt is made to first fully separate and cluster each shower and then determine the properties of the individual cluster. Furthermore, the set of steps, usually comprising at least one seeding step, and subsequent refinement of the object associated to the seed, all have similar pattern recognition requirements which can lead to suboptimal usage of resources.

An alternative to such a sequential reconstruction are machine-learning (ML) approaches based on the raw detector hit information, providing the possibility to reconstruct showers in parallel, and naturally accounting for overlapping energy deposits. Furthermore, ML algorithms are highly parallelisable, therefore, they exhibit a significant gain in execution speed on dedicated hardware such as graphics processing units (GPUs) or field programmable gate arrays~\cite{Duarte:2018ite,DiGuglielmo:2020eqx,Iiyama:2020wap,Heintz:2020soy,Aarrestad:2021zos}. In particular, recent developments on graph neural networks (GNNs)~\cite{qasim2019learning} have shown promising physics potential while keeping the computing resource requirements within expected financial and technological constraints~\cite{Aarrestad:2020ngo}.
Using GNNs, one can abstract from the detector geometry, which in general and in case of the HGCAL is irregular.

Reconstructing multiple particles without employing seeds is a challenging problem, in particular for ML algorithms aiming to also retain the information of nearby showers until the final properties are determined. However, the object condensation approach~\cite{kieseler2020object} has recently been proposed as a solution to this problem. Object condensation allows one to reconstruct multiple objects and their properties from irregular data in a one-shot approach by providing a method to quantify reconstruction efficiencies, noise suppression, and object property determination in a unified loss function.


In this contribution, we demonstrate the use of the GravNet graph neural network with an object condensation loss function to perform a seedless end-to-end clustering algorithm in the CMS HGCAL. We demonstrate that this technique can achieve the reconstruction performance needed in particle-dense environments at the HL-LHC while respecting computational constraints.  


\section{Dataset}
\label{sec-dat}
Dedicated simulated samples of particles interacting with the CMS detector are used to train our model and to evaluate the results. In order to demonstrate state-of-the-art performance in a complex environment that approaches that of an HL-LHC collision, we produce a data set with one hundred primary particles per event emanating from the interaction point at the center of the detector towards the acceptance regions encompassing the HGCAL in the CMS detector endcap. This leads to approximately 400 showers in the endcaps per event. The particle types are selected randomly per event from electrons, muons, photons, charged and neutral pions, and neutral, charged, short- and long-lived kaons. The energy and direction of the primary particles is selected randomly, evenly distributed between 20 and 200 GeV and in $\eta$, $\phi$, subject to the $\eta$ acceptance constraint. This leads to a falling energy spectrum of the particle showers as also shown in Figure~\ref{figefftrue}. The interaction of these particles with the detector and the consequent detector response are simulated through a detailed detector description in GEANT4~\cite{AGOSTINELLI2003250}. Simulated events have an average of $\approx$50,000 reconstructed hits (rechits) in the HGCAL detector, about one third of the average number expected in a t$\bar{t}$ event at 200 pileup. Our training data set comprises 32,000 collisions, which are divided by endcap to give a training data set of 64,000 entries. We evaluate performance with a test set of 1,250 collisions (2,500 entries).

\begin{figure}[h!]
  \centering
  \includegraphics[width=0.49\textwidth]{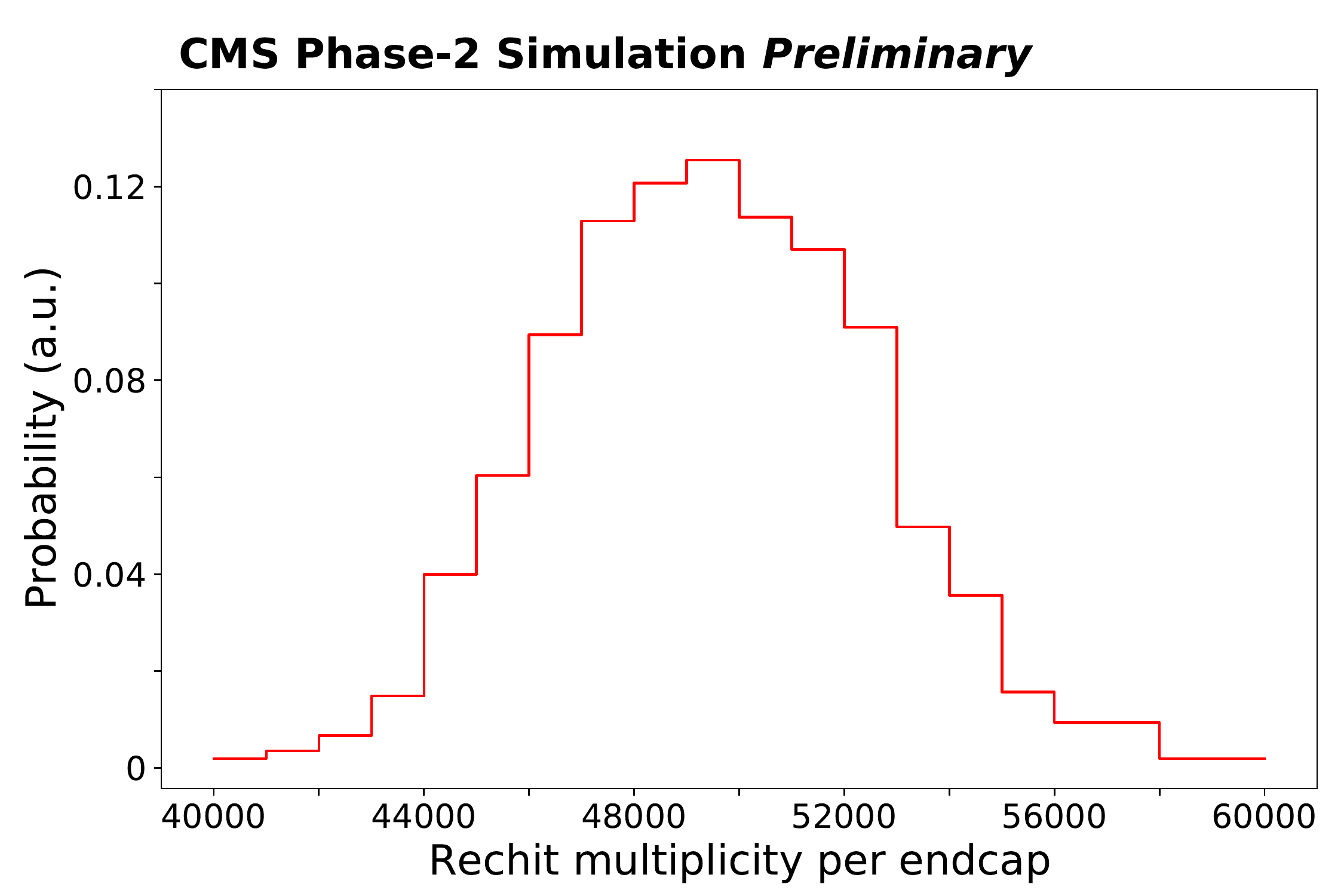}
  \vskip\baselineskip
  \includegraphics[width=0.49\textwidth]{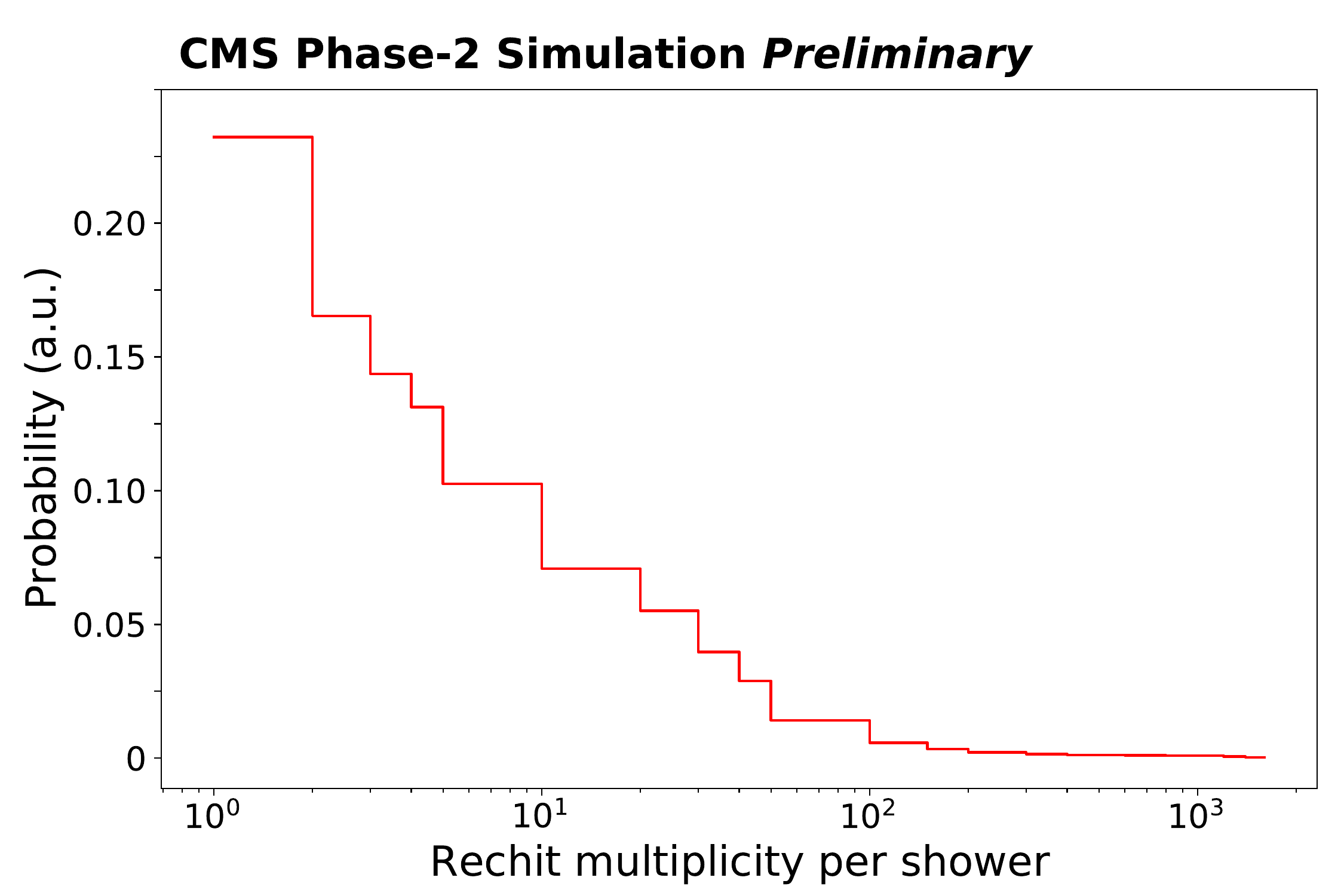}
  \includegraphics[width=0.49\textwidth]{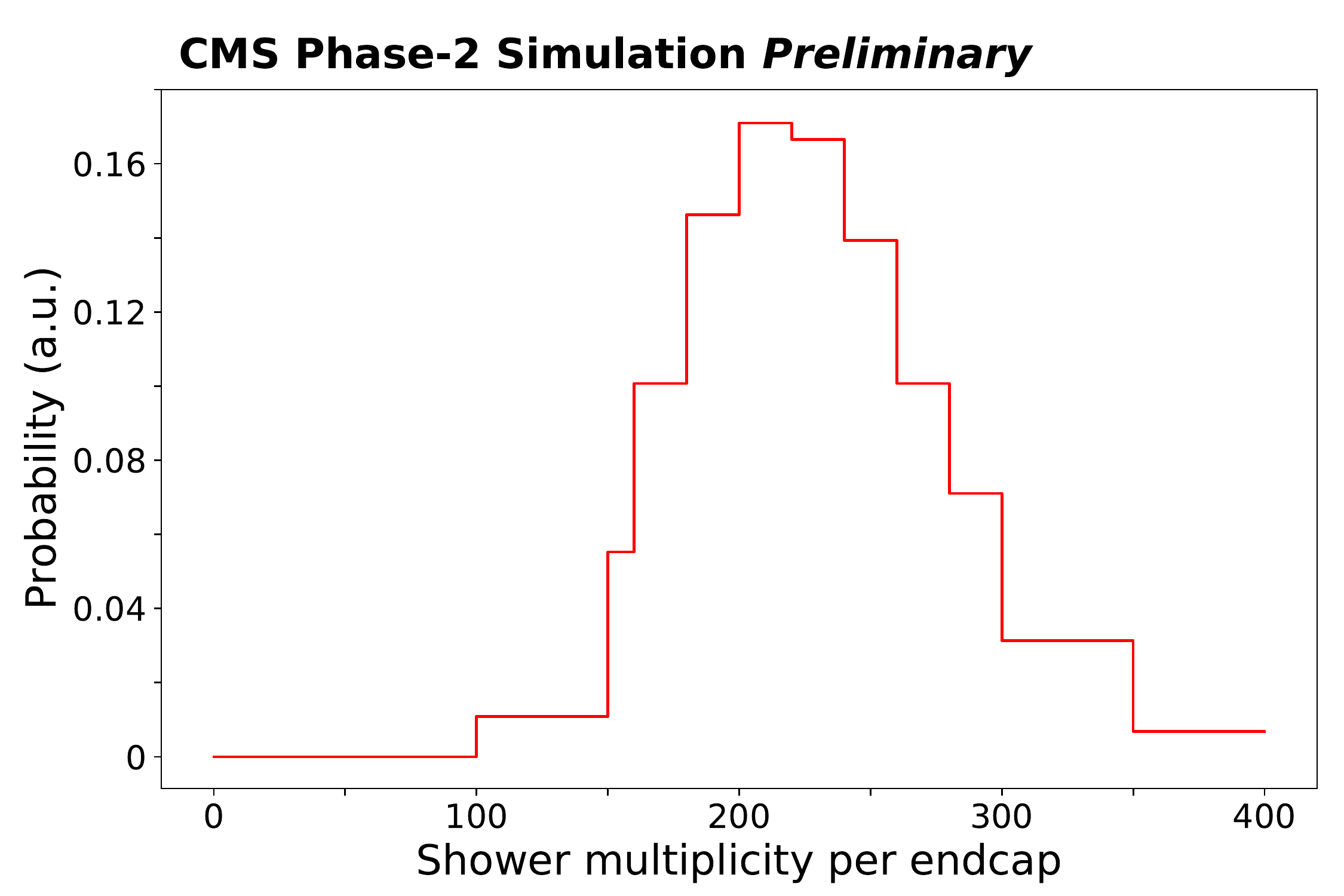}
  \caption{Properties of the reconstructed hits and true clusters for the simulated data set. The distributions of the number of rechits per endcap (top), the number of rechits (bottom left), and the number of true clusters per endcap (right) are shown. The distributions are normalized to unit area. Each endcap can contain up to 60,000 rechits and up to 400 showers. Each of these showers can have up to 1600 rechits.}
  \label{figdist}
\end{figure}

From the point of view of a machine-learning algorithm, the fact that the sensors of the HGCAL change shape and size throughout the detector leads to an irregular geometry that cannot be represented by a regular grid. Therefore, we represent each event as a set of sparse rechits (sensors with a signal over threshold), which are used as input into our reconstruction algorithm. Each rechit $r$ is associated with a feature vector of its reconstructed properties. These features comprise the spatial coordinates of the sensor recording the energy $\vec{x}_{r}$ in $xyz$ space, the time of detection $t_{r}$ and the recorded energy $e_{r}$.  Hence, all the $N$ rechits are defined as set $R = \{ r_i \mid r_i = \vec{x}(r) \oplus t(r) \oplus e(r) \}$\footnote{The symbol $\oplus$ represents concatenation}. The distribution of rechits arising from the primary and secondary interactions, as well as spurious hits due to noise in the detector, is shown in Figure~\ref{figdist}, alongside other general characteristics of the dataset.

Each rechit is additionally associated with a truth vector, representing the target properties of the clusters to which the rechits are associated. The true clusters are defined by exploiting the event history of the GEANT4 simulation. Distinct clusters are formed by collecting all energy deposits in the HGCAL volume that arise from an initiating particle that enters the HGCAL volume, or from secondary particles connected to the initiating particle in the GEANT4 event history, that is, daughter particles from conversions, radiations, or decays. If a decay or shower begins before crossing the plane of the HGCAL active detector volume, we consider the associated hits as separate clusters provided the separation of the two entry points $\Delta R = \sqrt{\Delta\phi^2+\Delta\eta^2} < 0.016$. When showers are merged, all energy deposits from associated particle interactions are collected and merged into a single shower. The set of all the truth showers is represented as $T = \{t_i \mid t_i \subseteq R\}$. Each of the truth shower is associated truth energy $E_{\mathrm{true}}(t)$ which is defined as the sum of the total deposited energy.

\section{Methodology}
\label{sec-meth}
Given the sparse and unordered nature of the input rechit sets, graph neural networks are a natural choice to perform end-to-end reconstruction tasks from raw data with deep neural networks~\cite{qasim2019learning}. 

The end-to-end reconstruction consists of multiple tasks: rechit clustering; particle identification (i.e., the association of a given cluster to any of the particle class, namely electron, muon, photon, or hadron); energy and position regression for each cluster. The methods discussed in this work allow these tasks to be performed and optimized simultaneously, however, we focus here on clustering and energy regression leaving identification tasks to future studies. The algorithm is optimized using the objection condensation loss\, which offers the possibility to handle inputs with variable length and to integrate regression and classification tasks via the minimisation of a common loss function.

\subsection{Architecture}
GravNet~\cite{qasim2019learning} is a distance weighted dynamic graph neural network which aims to build an extended representation of each vertex in the graph by aggregating information from those  vertices connected to it by the graph’s links. This results into a new representation of each vertex of the graph at the output of each layer, i.e., a new engineered feature vector. An individual GravNet layer transforms the input into two different spaces: feature space and low-dimensional spatial coordinates. For each vertex, the 64 nearest neighbors are determined in the spatial coordinates of each vertex. Then, the feature vector in the feature space of the neighbours are weighted by the edge length using a Gaussian function and gathered at each node using two aggregation functions: the mean and max of the distance-weighted features. Finally, the new vertex features are computed based on the original vertex features and the aggregated neighbor features by applying a learnable local transformation and tanh activation.
This architecture comes with significant advantages in terms of computing resources compared to other dynamic graph neural networks, and is therefore well suited for the challenging HGCAL data.

\begin{figure}[h!]
  \begin{center}
      \includegraphics[width=1\textwidth]{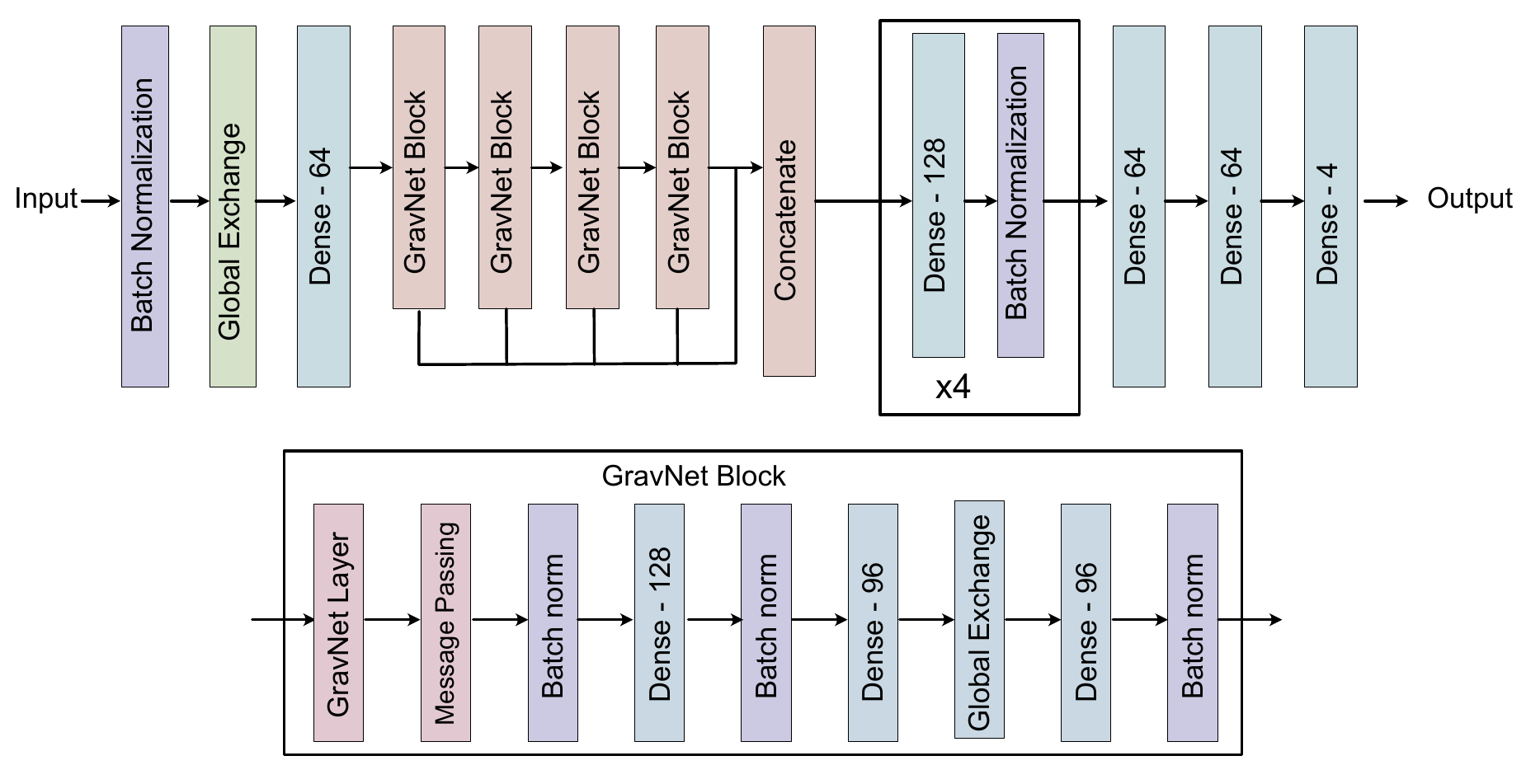}
  \end{center}
  \caption{Architecture diagram of the GravNet graph neural network.}
  \label{figarchitecture}
\end{figure}

The overall network architecture that we employ is schematically represented in Figure~\ref{figarchitecture}. The network consists of four GravNet blocks, each extended with multiple neural network layers.  After each GravNet layer, there are 6 message passing layers in each GravNet block which employ the same accumulation function without distance weighting. The resulting features of GravNet and all message passing layers are concatenated into a feature vector for each vertex. 

\subsection{Training}

The object condensation approach aims to accumulate all object properties in \textit{condensation points}. Out of all vertices, the condensation points are identified by a learnable confidence measure $\beta$, where high values of $\beta$ indicate a more significant condensation point. Condensation points attract the other points belonging to the same object and repulse noise or points from other objects in a latent clustering space. The object properties, in this case the shower energy, are also predicted per point and weighted by $\beta$, such that the cluster centers also embed the best predictions for the cluster properties. Hence, the output of our graph neural network consists of 4 features per node in total: the clustering space $C \in \mathbb{R}^{2}$, the $\beta$ confidence, and $E_{\mathrm{pred}}$. The loss function consists of three terms:

\[ L = a L_V + b L_{\beta} + c L_{P} \]

$L_V$ describes the attractive and repulsive potential losses as defined in Ref.~\cite{kieseler2020object}.
In the present work, we take $a=1$, $b=10$, and $c=1$. The attractive potential is given by a second order polynomial and acts from a representative vertex on all vertices belonging to the same object. Similarly, a linear repulsive term repulses vertices belonging to other objects from the representative vertex.

For every truth shower $t$, the vertex which has the highest $\beta$ confidence value is chosen as the representative vertex ($\alpha(t)$).

\[ \alpha(t) = \mathrm{argmax}(\beta(t)) \]

The beta loss ($L_{\beta}$) is used to maximise the $\beta$ confidence value for the representative vertex \cite{kieseler2020object}. The term $L_{P}$ stands for payload loss. It is used to minimize the error in property prediction for every cluster which, in our case, only comprises the energy of the shower and is defined below:

\[ L_P = \frac{1}{|T|} \sum_{t \in T}{} \frac{1}{\sum_{r \in t}^{} \xi(r)} \sum_{r \in t}^{} (L_E(E_{\mathrm{true}}(r), E_{\mathrm{pred}}(r)) \xi(r) \]

where
\[ \xi(r) = \arctanh^2(\beta(r)) \],

and $L_E$ is the energy loss. In principle, this term can take a simple form such as a mean squared error. However, to account for the expected resolution of the calorimeter as well as possible outliers, we take

\[L_E(E_{\mathrm{true}}(r), E_{\mathrm{pred}}(r)) = L_\delta( \frac{E_{\mathrm{true}}(r) - E_{\mathrm{pred}}(r)}{\sqrt{E_{\mathrm{true}}(r)} + 1}, 2\sqrt{E_{\mathrm{true}}(r)})\]
$L_\delta$ is the Huber loss\cite{Huber1964RobustEO}:

\[ L_\delta(\Delta,\theta) =
\begin{cases}
    \Delta^2, & \text{if } \abs{\Delta}\leq \theta\\
    \theta^2 + 2\theta(\abs{\Delta} - \theta) ,& \text{otherwise}
\end{cases}
\]

We employ a cyclic learning rate scheduling~\cite{smith2017cyclical} and optimise the weights using the Adam optimizer~\cite{kingma2014adam} with Nesterov momentum~\cite{nesterov_momentum,Dozat2016IncorporatingNM}.

\subsection{Inference}
As described in Ref.~\cite{kieseler2020object}, we select only points above a threshold $\beta_{\mathrm{thresh}}$. Then we start with the highest $\beta$ point, assign everything within a $d_{\mathrm{thresh}}$ radius in the clustering space as predicted shower $p$. The highest $\beta$ point is taken as the representative vertex ($\alpha(p)$). The reconstructed energy for the shower $p$ is taken from the representative vertex, such that
\[ E_{pred}(p) = E_{\mathrm{pred}}(\underset{\beta}{\mathrm{argmax}}(p))\].

This process repeats until we have depleted all the points above $\beta_{\mathrm{thresh}}$, collecting all the predicted showers $P$. Everything else is considered noise. Optimal values of $\beta_{\mathrm{thresh}}$ and $d_{\mathrm{thresh}}$ are selected after training by optimizing the efficiency for true shower identification against the rate of clusters not matched to a true shower (fake rate). A low value of $\beta_{\mathrm{thresh}}$ and $d_{\mathrm{thresh}}$ will correspond to more showers being reconstructed, leading to higher efficiency, but also a higher fake rate. We find that the optimal value for both parameters is $0.6$.

\begin{figure}[h!]
  \centering
  \includegraphics[width=0.49\textwidth]{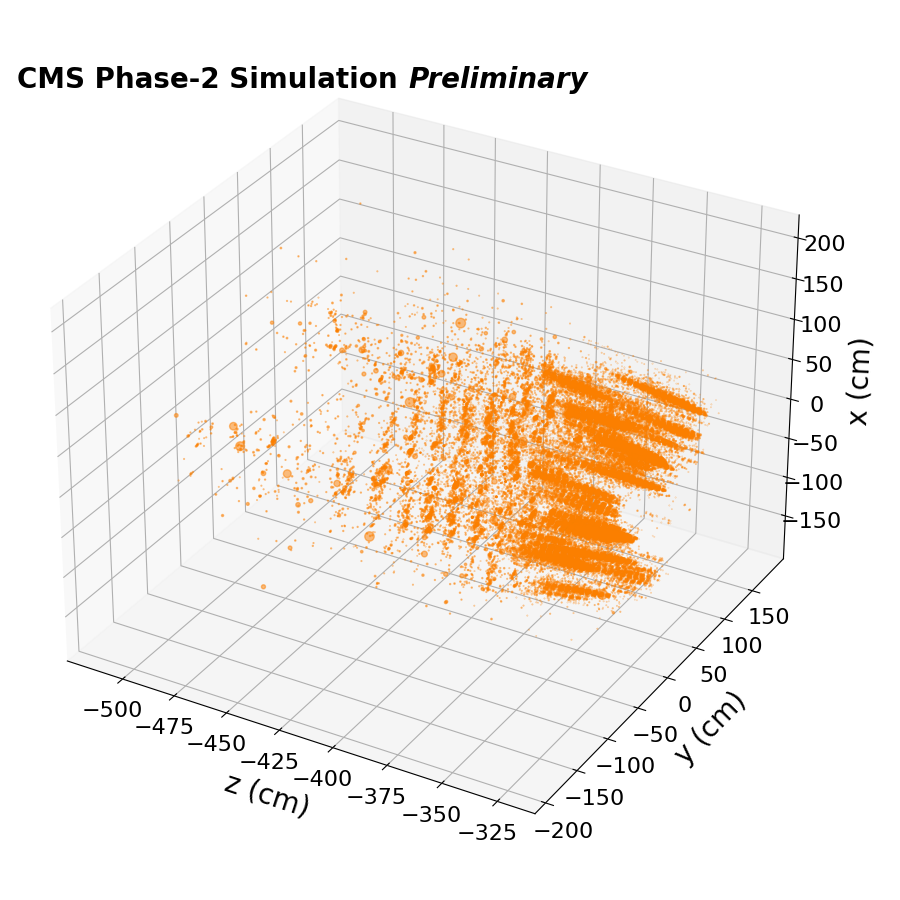}
  \vskip\baselineskip
  \includegraphics[width=0.49\textwidth]{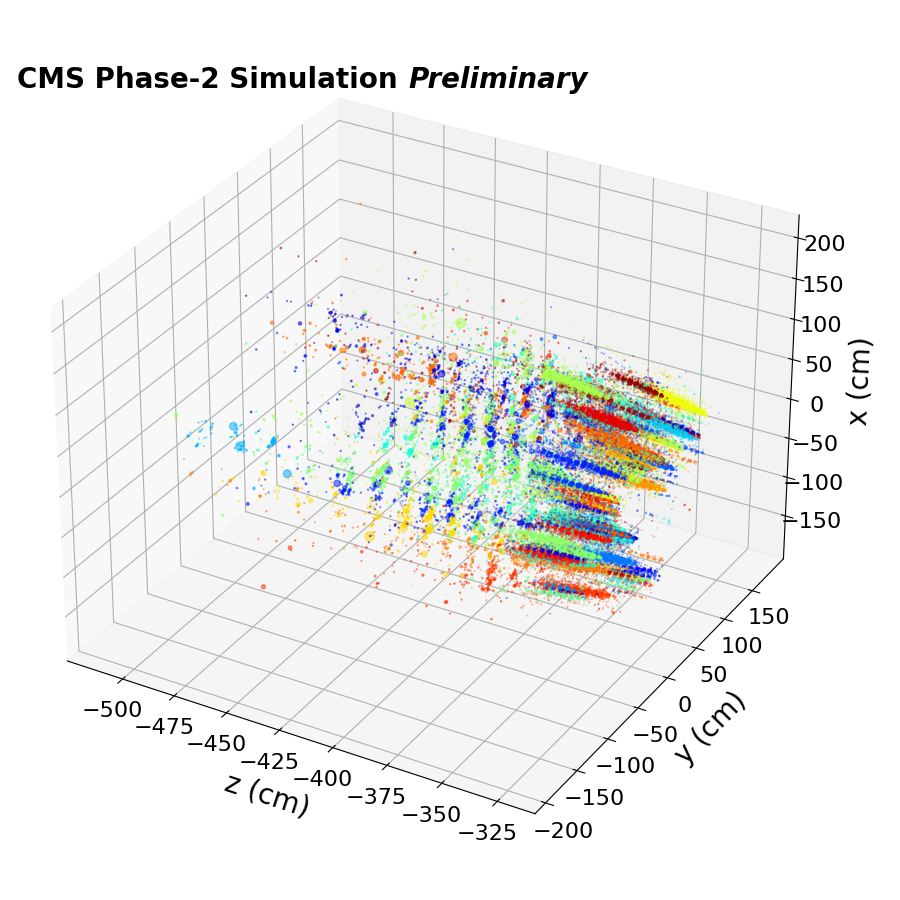}
  \includegraphics[width=0.49\textwidth]{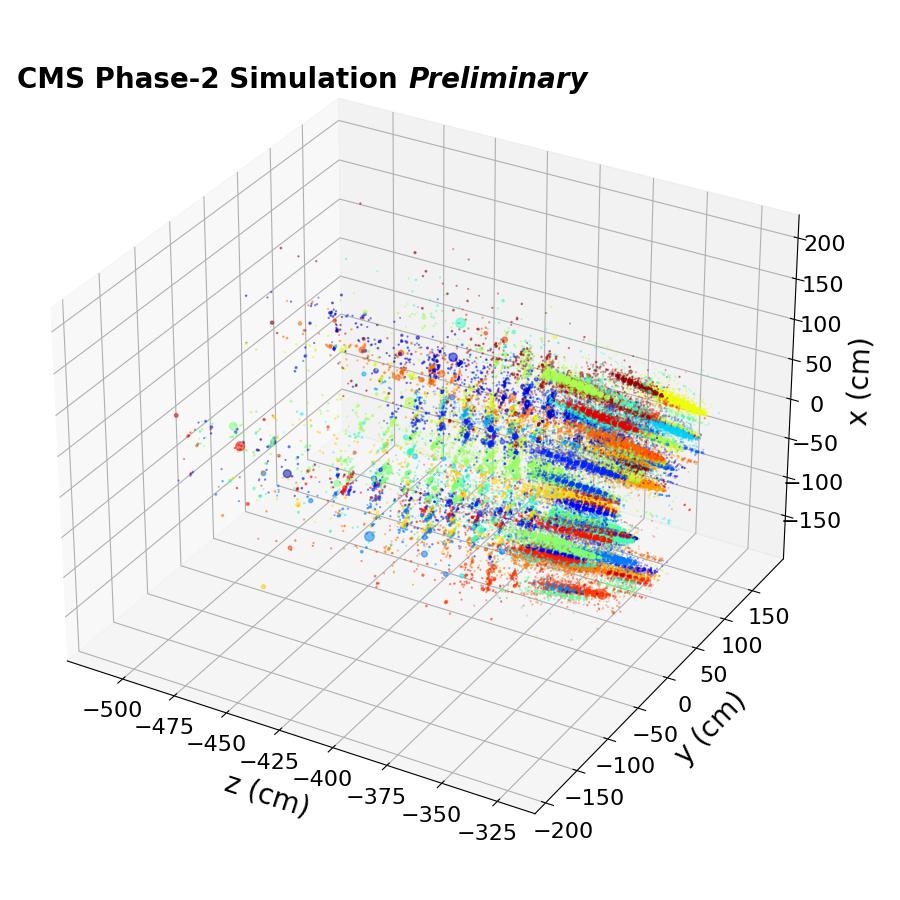}

  \caption{The top plot shows the input to the network. Different colors represent true clusters in the bottom left plot and reconstructed clusters in the bottom right plot.}
  \label{figsample}
\end{figure}

A sample of network input, output, and ground truth is shown in Figure~\ref{figsample}.



  
  
    
    
    


\begin{figure}
     \centering
     \begin{subfigure}[b]{0.475\textwidth}
         \centering
         \includegraphics[width=\textwidth]{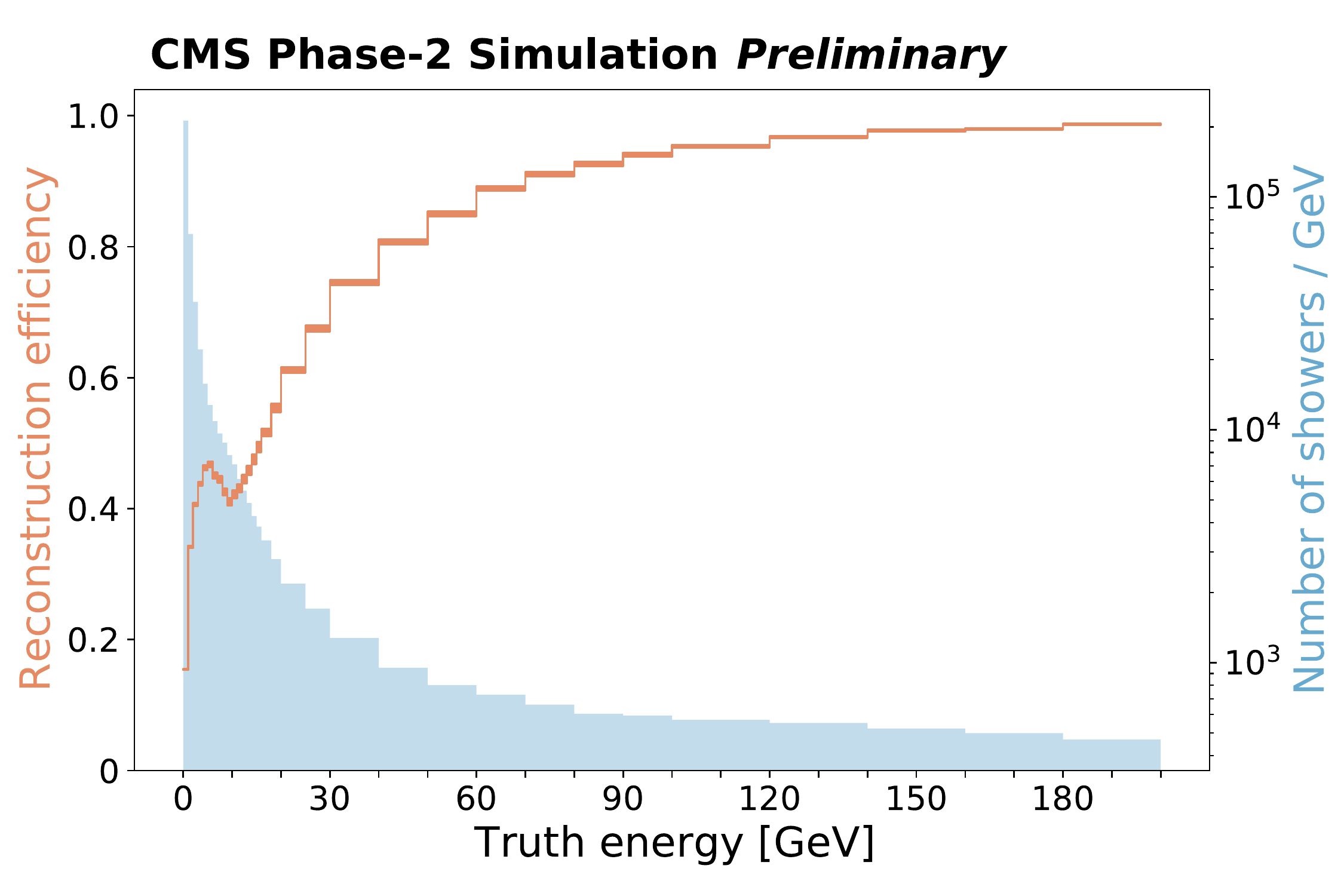}
         \caption{The reconstruction efficiency as a function of the true cluster energy}
         \label{figefftrue}
     \end{subfigure}
     \hfill
     \begin{subfigure}[b]{0.475\textwidth}
         \centering
         \includegraphics[width=\textwidth]{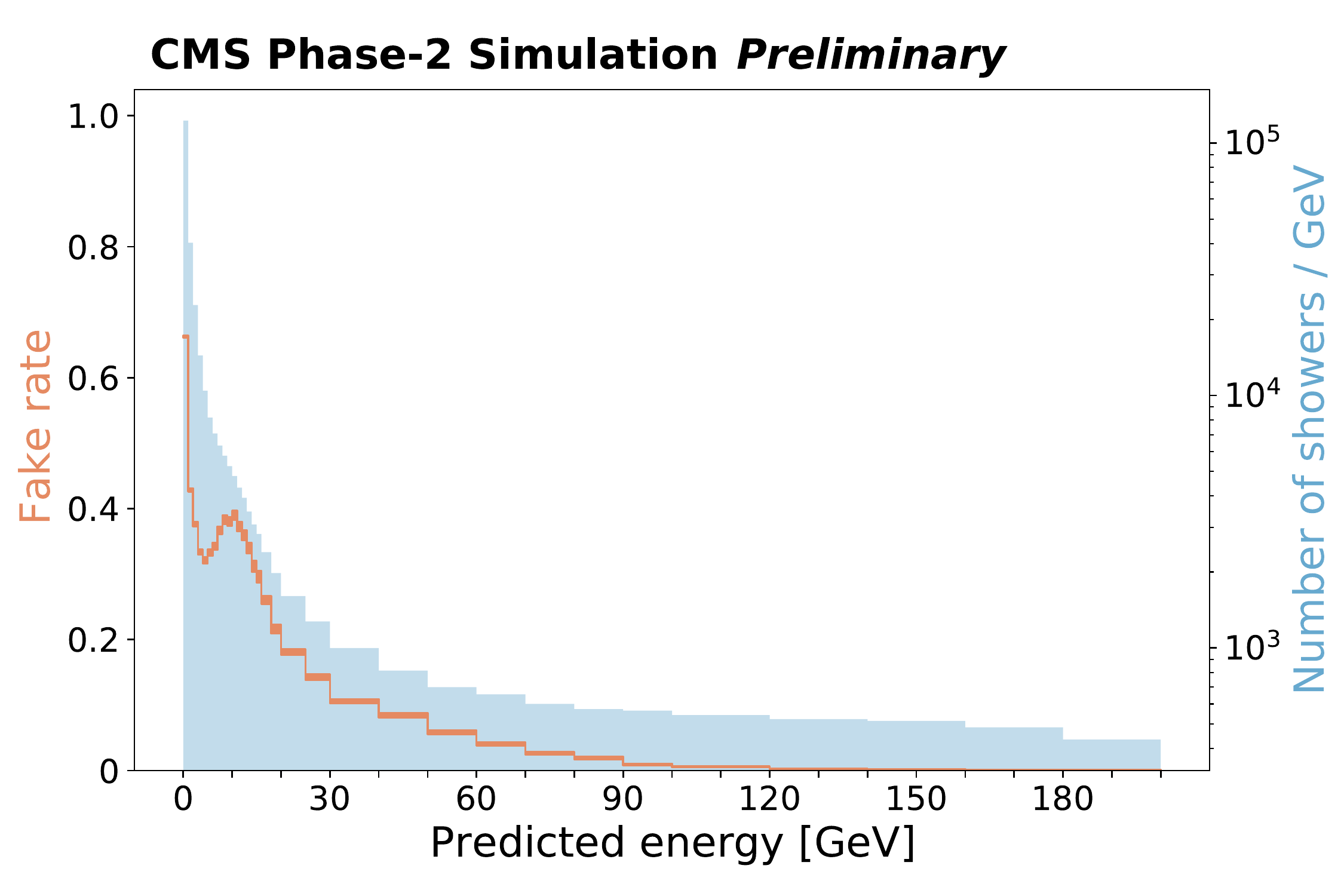}
         \caption{The fake rate as a function of the reconstructed cluster energy}
         \label{figfakepred}
     \end{subfigure}
     \vskip
     \baselineskip
     \begin{subfigure}[b]{0.475\textwidth}
         \centering
         \includegraphics[width=\textwidth]{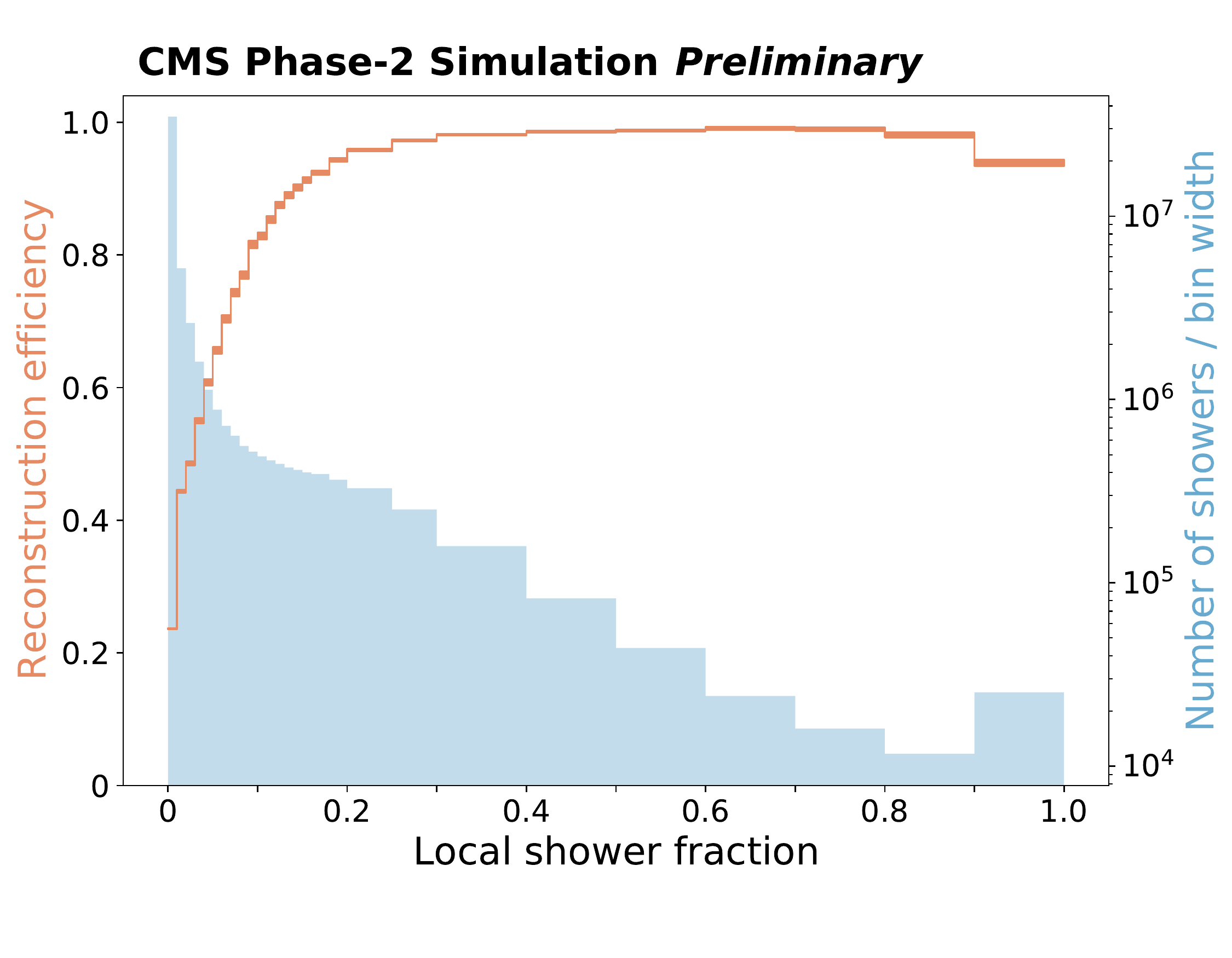}
         \caption{The reconstruction efficiency as a function of local shower fraction}
         \label{figefflocal}
     \end{subfigure}
     \hfill
     \begin{subfigure}[b]{0.475\textwidth}
         \centering
         \includegraphics[width=\textwidth]{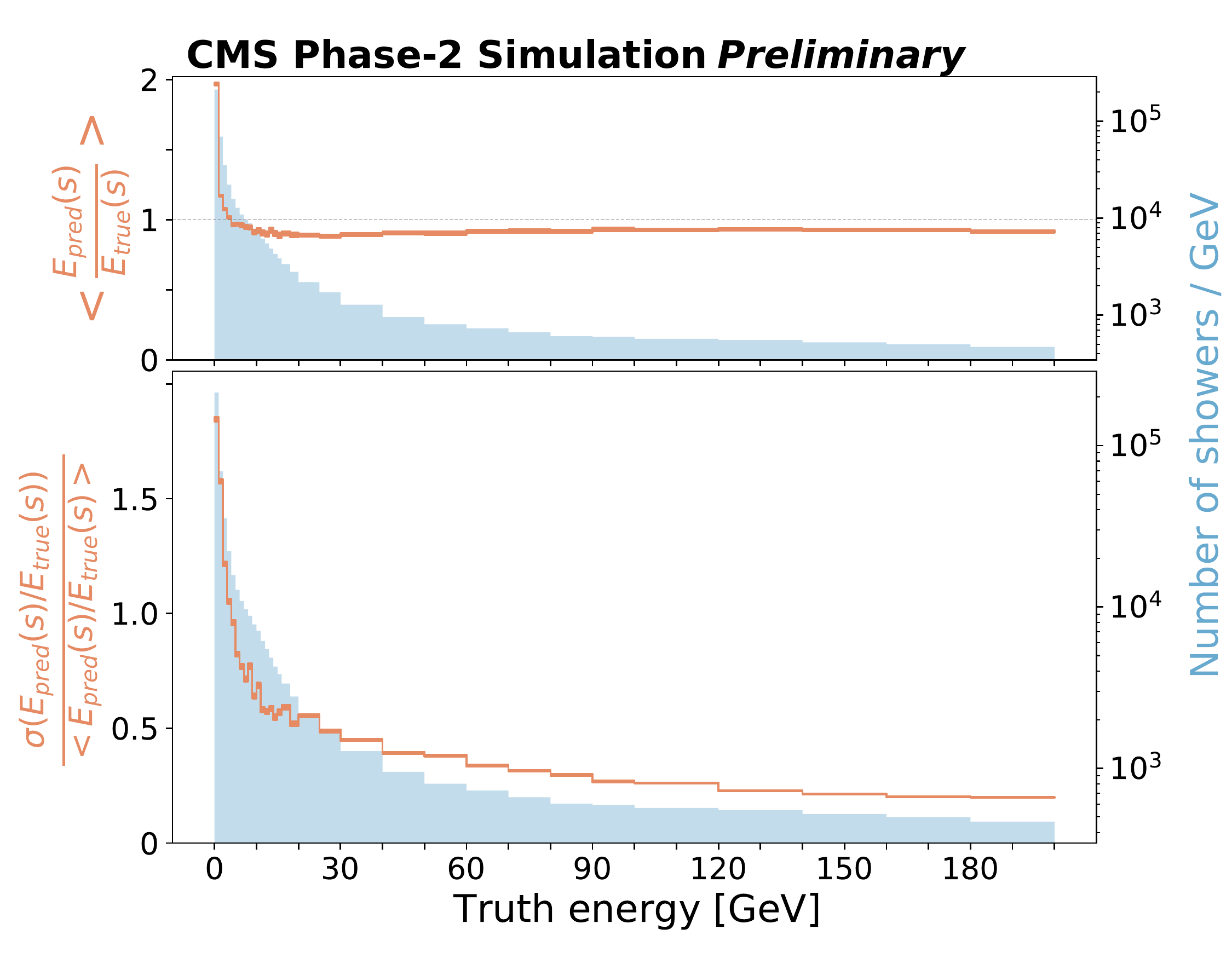}
         \caption{The energy response and energy resolution as a function of true cluster energy}
         \label{figrestrue}
     \end{subfigure}

    \caption{Quantitative metrics illustrating the performance of our reconstruction algorithm. The distributions in orange correspond to the observable indicated in the y-axis to the left. The blue histograms show underlying distributions of x-axis variables with respect to the observable indicated on the y-axis to the right. The line width of the orange distribution corresponds to the statistical error.}
    \label{figquant}
\end{figure}

\section{Analysis and results}
\label{sec-analy}
Because clustering is a complex task involving the association of many objects, evaluating the performance of a clustering algorithm is not as straightforward as for tasks such as single-object classification. To build an appropriate metric to quantify the reconstruction performance, we take inspiration from the task of object detection in the field of computer vision, where the metric intersection over union (IOU) or Jaccard Index \cite{jaccard1901etude} is used. The IOU is defined as the number of pixels in the intersection of two objects divided by the number of pixels in the union. However, this definition implicitly assumes that all pixels carry equivalent importance, which is not the case for our reconstruction task, where rechits with higher energy deposits carry greater importance in our clusters of reconstructed showers. Therefore, we extend this metric to define an energy weighted intersection over union (EIOU), defined as

\[ \mathrm{EIOU}(t,p) =  \frac{\sum_{i}^{} e(r_i) \mid r_i \in (t \cap p)}{\sum_{i}^{} e(r_i) \mid r_i \in (t \cup p)}\].

The EIOU is calculated for each pair of ground truth and predicted shower, i.e., $\forall \{P \times T\}$. In order to exclude very small overlaps from the matching procedure, we require EIOU $>$ 0.1.

In a next step, predicted clusters are matched to the true shower clusters. A simple algorithm in which a ground truth shower is matched to the predicted shower with the highest overlap or vice versa is not optimal; in particular, such an approach depends on the order in which the association is made. 
Instead, we aim to optimise the matching of all showers simultaneously. This can be framed as an assignment problem, namely

\[ M = \mathrm{argmax}(\sum_{}^{}{EIOU (x)}) \forall x \in \mathbb{C}(\mathbb{P}(\{P \times T\}))\].

Here, $\mathbb{P}$ takes the power set defining the domain and $\mathbb{C}$ defines two constraints: a) each true cluster can match to only one predicted shower and vice versa, and b) Each matching cluster should have a minimum EIOU of $0.1$. In order to perform this matching efficiently, we use the algorithm of Ref.~\cite{galil1986efficient}

We denote reconstructed clusters that are matched to a true cluster as ``matched clusters'', and all reconstructed clusters which are not matched to a true cluster as ``fake clusters''. We define the reconstruction efficiency as the number of matched clusters divided by the number of true clusters. The fake rate is defined as the number of fake clusters divided by the total number of reconstructed clusters. The performance of our algorithm with respect to these metrics is shown in Figure~\ref{figefftrue} and Figure~\ref{figfakepred} as a function of the true cluster energy. The reconstruction efficiency increases with energy, but shows reasonable performance already at very low energies, a region which will crucially determine the total physics performance, allowing to either remove pileup contributions from jets, or determine a precise missing transverse momentum. Similarly, the fake rate is also higher at lower energies  because of a high level of stochasticity and the similarity of low energy showers and noise. Moreover, this fake rate does not necessarily pose a problem since it is possible that the neural network decided to split one truth shower into two reconstructed showers, out of which one would be identified as ``fake''. In this case, the energy weighted truth matching algorithm would consider the higher energy shower as matched, which also increases the fake rate at low energies. The decrease in efficiency and increase in fake rate at 10$\,$GeV is caused by the transition region between primary particles entering the HGCAL (above approximately 20 GeV) and only secondary particles being present (below 20 GeV). Therefore, it is an artifact of the approach we use to generate our data set and not a feature of our reconstruction algorithm---the exact same reduction in efficiency is observed when employing purely human-engineered algorithms. In addition, the region with only secondary particles, the truth information in the data set is not fully reliable due to known insufficiencies of the simulation.


The reconstruction efficiency also depends on the cluster energy. As a metric to describe the relative energy density around the shower cluster in question, we define the local shower fraction (LSF) as the fraction of energy belonging to a given cluster with respect to the total energy within a radius of $\Delta R=0.5$ center point of the cluster. That is,


\[\mathrm{LSF}(s) = \frac{E_{true}(s)}{\sum_{i}^{} E_{true}(i) \mid \Delta R(s, i) < 0.5 }\text{.}\]
The dependence of the reconstruction efficiency on the LSF is shown in Figure~\ref{figefflocal}. The efficiency increases with the LSF, reflecting the fact that more spatially concentrated showers are easier to reconstruct. However, even for low LSF values of $\sim$10\%, the reconstruction efficiencies approaches 70\%. Because the low LSF region is important for the removal of pileup particles from jets and for jet substructure---for example, in order to separate quark and gluon jets in VBS or VBF event---efficient reconstruction in this region is an important result.

Finally, the reconstruction efficiency and energy resolution are shown in Figure~\ref{figrestrue}. The network has a tendency to overestimate the energy of low energy clusters, and a slight tendency to underestimate the energy of high-energy clusters.
However, the energy response and resolution are promising, given the fact that this is the first application of simultaneous clustering and property determination to a high dimension reconstruction task. We expect future work to exceed this benchmark performance.

\section{Conclusions}
\label{sec-conc}
This work describes the first application of dedicated graph neural networks for simultaneously separating, clustering and determining the properties of individual showers from all hits in a full subdetector. We show that these tasks can be performed even in a highly dense environment and a highly-granular calorimeter such as the CMS HGCAL, with up to 60$\,$000 hits in a single step without the need for seeding. Using object condensation, property determination and clustering can also be optimized simultaneously.
While we focus on estimating energy of clustered showers here, the identification of the incident particles and other properties are natural extensions of this work. Finally, we propose a technique to match truth and reconstructed clusters by maximising the energy weighted intersection over union, which naturally generalizes to other reconstruction approaches and will become more important with a more dense environment. Using this advanced matching algorithm, 
we demonstrated the potential for end-to-end multi-particle reconstruction algorithms based on graph neural networks in the highly challenging environment of HL-LHC collisions.

\section*{Acknowledgements}
This project has received funding from the European Research Council (ERC) under the European Union's Horizon 2020 research and innovation program (Grant Agreement No. 772369)

\bibliography{main}{}

\end{document}